\documentclass[10pt]{article}
\usepackage{graphicx}

\def\Title#1{\begin{center} {\Large #1 } \end{center}}
\def\Author#1{\begin{center}{ \sc #1} \end{center}}
\def\Address#1{\begin{center}{ \it #1} \end{center}}

\newcommand\pubblock{\rightline{\begin{tabular}{l} Proceedings of the Fifth Annual LHCP\\ 
         \pubdate  \end{tabular}}}

\newenvironment{Abstract}{\begin{quotation} \begin{center} 
             \large ABSTRACT \end{center}\bigskip 
      \begin{center}\begin{large}}{\end{large}\end{center} \end{quotation}}

\newenvironment{Presented}{\begin{quotation} \begin{center} 
             PRESENTED AT\end{center}\bigskip 
      \begin{center}\begin{large}}{\end{large}\end{center} \end{quotation}}

\def\Acknowledgements{\bigskip  \bigskip \begin{center} \begin{large}
             \bf ACKNOWLEDGEMENTS \end{large}\end{center}}

\def\beq{\begin{equation}}
\def\eeq#1{\label{#1}\end{equation}}
\def\eeqn{\end{equation}}

\def\beqa{\begin{eqnarray}}
\def\eeqa#1{\label{#1}\end{eqnarray}}
\def\eeqan{\end{eqnarray}}

\let\bar=\overbar

\def\Dslash{\not{\hbox{\kern-4pt $D$}}}
\def\dslash{\not{\hbox{\kern-2pt $\del$}}}

\def\msb{{\bar{\ssstyle M \kern -1pt S}}}

\usepackage[svgnames]{xcolor} 
\usepackage{slashed}

\newcommand{\bea}{\begin{eqnarray*}}
\newcommand{\eea}{\end{eqnarray*}\nolinebreak}
\newcommand{\nn}{\nonumber\\}

\newcommand\pubdate{August 22, 2017}

\def\affiliation{Faculty of Physics, University of Warsaw, \\
Pasteura 5,  02-093 Warsaw, Poland}

\begin{document}

\large
\begin{titlepage}
\pubblock

\vfill
\Title{GENERAL MASS INSERTION EXPANSION IN FLAVOR PHYSICS}
\vfill

\Author{ JANUSZ ROSIEK  }
\Address{\affiliation}
\vfill

\begin{Abstract}
Calculating amplitudes for the flavor changing transitions in terms of
the off-diagonal elements of mass matrices, so called “Mass
Insertions” (MI) in the theory defined in “gauge basis” (before mass
matrix diagonalization) is the common technique in analyzing the
flavor structure of the New Physics models. I present a general method
allowing to expand any QFT amplitude calculated in the
mass-eigenstates (physical) basis into series in MI's, to any required
order~\cite{Dedes:2015twa} . The technique is purely algebraic,
translating an amplitude written in the mass eigenbasis into MI series
without performing diagrammatic calculations in the gauge basis. It
can be applied for all types of mass matrices - either Hermitian
(scalar or vector), general complex (Dirac fermions) or complex
symmetric (Majorana fermions). Proposed expansion has been also
automatized in the form of publicly available specialized Mathematica
package, {\em MassToMI}~\cite{Rosiek:2015jua}, which features I
illustrate with the example of the Higgs boson decays in the MSSM.
\end{Abstract}

\vfill

\begin{Presented}
The Fifth Annual Conference\\
 on Large Hadron Collider Physics \\
Shanghai Jiao Tong University, Shanghai, China\\ 
May 15-20, 2017
\end{Presented}
\vfill
\end{titlepage}
\def\thefootnote{\fnsymbol{footnote}}
\setcounter{footnote}{0}
%

\normalsize 

\section{Introduction:  basis choice for flavor transition calculations}

\noindent Amplitudes describing flavor changing transitions within the
New Physics models can be computed using two basic approaches:
\begin{itemize}
\item Calculations in the ``interaction'' or ``symmetry'' basis, using
  fields before the mass matrices diagonalization.  Gauge interactions
  in such basis are flavor diagonal, flavor transitions originate from
  the off-diagonal entries of mass matrices, so-called ``Mass
  Insertions'' (MI).
  \begin{itemize}
  \item advantage: direct dependence on original symmetry-related
    parameters.
  \item disadvantage: tedious and error prone diagrammatic
    calculations with mass insertions treated as interaction vertices,
    combinatorial complication quickly growing with MI order.
    Amplitude expressed as double infinite series, in loop order and
    in MI order.
  \end{itemize}
\item Calculations in the ``mass eigenstates'' basis, in terms of
  physical fields after mass matrices diagonalization.
  \begin{itemize}
  \item advantage: physical external states, more compact expressions,
    easier diagrammatic calculations, at given loop order exact
    formulae in terms of flavor changing parameters.
  \item disadvantage: complicated non-linear dependence on initial
    symmetry-related parameters, various effects can be analyzed only
    numerically.
  \end{itemize}
\end{itemize}
Both approaches can be illustrated with the following toy example. Let
us consider the scalar self-energy in a model with one real scalar
field {$\eta$} and one complex scalar multiplet {$\Phi_I$}, defined by
the weak-basis Lagrangian:
\bea
L_{\rm int} = (\partial^\mu \Phi_I^\dag )\,(\partial_\mu \Phi_I)
-M_{IJ}^2 \Phi_I^\dag \Phi_J + \,\frac{1}{2}(\partial^{\mu}\eta) \,
(\partial_{\mu}\eta) - \frac{1}{2} m_\eta^2 \eta^2 \, - Y_{IJ}\;\eta\,
\Phi_I^\dag\, \Phi_J
\label{Luflavour}
\eea
Transition to mass eigenstates basis is done by the unitary rotation
${\bf U}$:
\bea
\Phi_I = U_{Ii} \phi_i \qquad
{\bf m^2} = \mathbf{diag} (m_1^2,\ldots, m_N^2) = {\bf U^\dagger \,
  M^2 \, U} \qquad
{\bf y} &=& {\bf U^{\dagger} \, Y \, U}
\eea
Then, self-energy amplitude in the interaction basis reads as (thick
dots in the diagram represent off diagonal elements of ${\bf M^2}$ and
$B_0,C_0,D_0,\ldots$ are the standard scalar $2-,3-,4-,\ldots$ point
1-loop functions):
%
\begin{center}
\includegraphics[trim=0.3cm 5.4cm 0cm 4cm, clip=true,
    totalheight=0.12\textwidth]{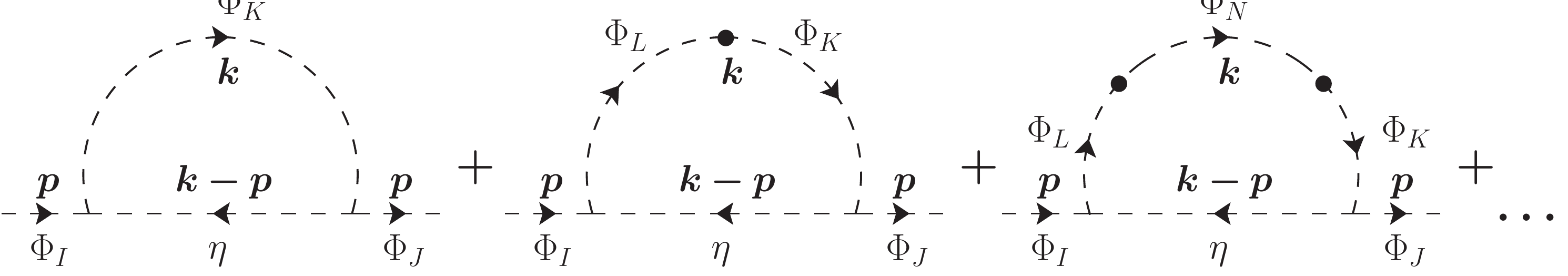} 
\end{center}
%
\bea
\hat{\Sigma}_{JI}(p)
& = & -\frac{1}{(4\pi^2)} Y_{JK} Y_{LI} \Bigg(\delta_{KL}
B_0(p;M_K^2,m_\eta^2)
+ \hat{M}_{KL}^2 C_0(0,p;M_K^2,M_L^2,m_\eta^2)\nn
&+& \hat{M}_{KN}^2\hat{M}_{NL}^2 D_0(0,0,p;M_K^2,M_N^2,M_L^2,m_\eta^2)
\ +\ \ldots\Bigg)\;,
\label{eq:miase}
\eea
while the mass eigenstates basis amplitude has the form:\\[4mm]
\begin{tabular}{lr}
\includegraphics[trim=0.5cm 4.cm 2cm 3.7cm, clip=true,
  totalheight=0.12\textwidth]{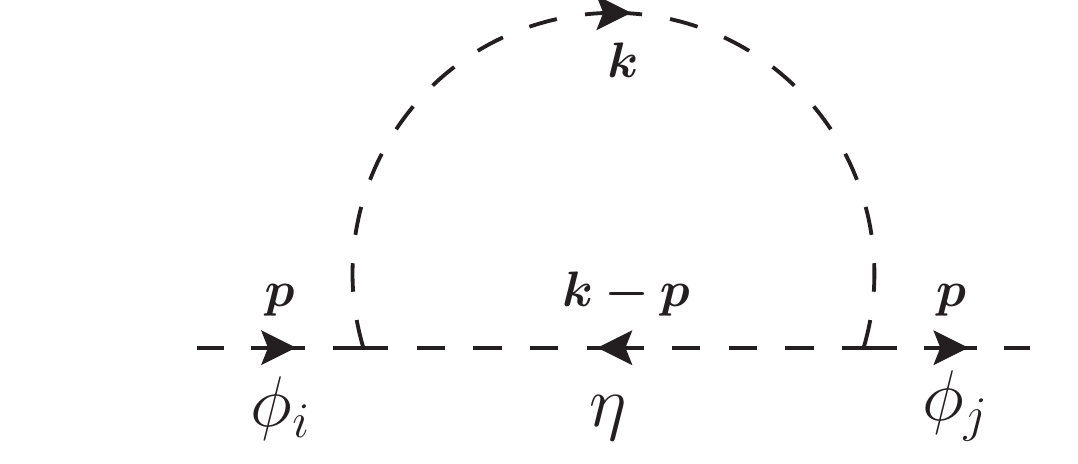}
& \begin{minipage}[b][10mm][t]{0.8\linewidth}{$\qquad\Sigma_{ji}(p) =
      U^\dag_{jJ} \hat{\Sigma}_{JI}(p) U_{Ii} = -\frac{1}{(4\pi)^2} \:
      y_{j\ell} \: B_0(p;m_\ell^2,m_\eta^2) \: y_{\ell i}
      $}\end{minipage}
\end{tabular}



\section{Mass Insertion expansion of bosonic amplitudes}

As proven in ref.~\cite{Dedes:2015twa}, transition from mass to
interaction basis amplitudes can be done using purely algebraic
techniques, without the diagrammatic calculations with Mass
Insertions.

Typical term in QFT mass-eigenstates amplitude involving bosonic
fields can be, through the Taylor series expansion, expressed as an
element of function of the Hermitian mass matrix $\mathbf{M^2}$:
\bea
U_{Ii} f(m^2_i) U_{Ji}^* & = & U_{Ii} \left(f_0 + f_1 m_i^2 + f_2
m_i^4 + \ldots \right) U_{Ji}^*
= (f_0 + f_1 \mathbf{M^2}+ f_2 \mathbf{M^4} + \ldots)_{IJ}
 = f(\mathbf{M^2})_{IJ}
\eea
MI approximation requires expanding $f(\mathbf{M^2})_{IJ}$ into a
series with following properties:
\begin{itemize}
\item \textit{Series coefficients:} depend only of diagonal elements
  of mass matrices.
\item \textit{Expansion parameters:} only non-diagonal elements of
  mass matrices.
\end{itemize}    
The relevant solution is given by the ``Flavor Expansion Theorem''
(FET), valid for any function of Hermitian matrix (assuming only
proper convergence criteria, see~\cite{Dedes:2015twa} for discussion
of the most important case of the Passarino-Veltman 1-loop functions).
The FET can be formulated as follows.  Let's decompose a Hermitian
matrix {$\bf M^2 $} as a sum of diagonal and non-diagonal part:
\bea
\mathbf{M^2} &=& \mathbf{M^2_0} +\mathbf{\hat{M}^2} \qquad
\mathrm{where} \qquad
\begin{array}{l}
(M^2_0)_I \equiv M^2_{II}\;\qquad \hat{M}^2_{IJ}\,\equiv\,M^2_{IJ}
  \qquad\hat{M}^2_{II}=0.
\end{array}\label{Adiag}
\eea
Then, matrix element {$ f({\bf M^2})_{IJ}$} is given by (no sum over
{$I,J$}) (see ref.~\cite{Dedes:2015twa} for formal proof):\\[-4mm]
\bea
f({\bf M^2})_{IJ} &=& \delta_{IJ}f((M^2_0)_I)\ +
f^{[1]}((M^2_0)_I,(M^2_0)_J)\:\hat{M}^2_{IJ}
+ \sum_{K_1} f^{[2]}((M^2_0)_I,(M^2_0)_J,(M^2_0)_{K_1})\:
\hat{M}^2_{I{K_1}}\hat{M}^2_{{K_1}J}\nonumber\\
& +& \sum_{K_1,K_2}f^{[3]}((M^2_0)_I,(M^2_0)_J,(M^2_0)_{K_1},(M^2_0)_{K_2})\:
\hat{M}^2_{I{K_1}}\hat{M}^2_{{K_1}{K_2}}\hat{M}^2_{{K_2}J} + \dots 
\label{theorem}
\eea
and $f^{[n]}$ are divided differences of function $f$, defined
recursively as:
\bea
f^{[0]}(x)&\equiv f(x)\nn
f^{[1]}(x_0,x_1)&\equiv\frac{f(x_0)-f(x_1)}{x_0-x_1}\nn
\ldots & \nonumber\\
f^{[k+1]}(x_0,\dots ,x_k,x_{k+1})&\equiv \frac{f^{[k]}(x_0,\dots ,
  ,x_{k-1},x_k)-f^{[k]}(x_0,\dots ,x_{k-1},x_{k+1})}{x_k-x_{k+1}}\nonumber
\eea
Using FET, any amplitude involving scalar or vector fields can be
calculated in mass eigenstates basis and consistently expanded to any
order in MI's (see e.g. ref~\cite{Dedes:2014asa} for example of
applications).

\section{Mass insertion expansion of fermionic amplitudes}

FET theorem holds for Hermitian matrices.  Mass terms for chiral
fermions depend on {{\em general complex}} mass matrix {$\mathbf{M}$}:
\bea
-\mathbf{\bar\Psi}\Big( \mathbf{M} P_L + \mathbf{M^\dag} P_R\Big)
\mathbf{\Psi}\;,
\eea
which can be diagonalized by 2 independent unitary rotations
{$\mathbf{U},\mathbf{V}$}, applied to left and right part of the
spinor field:
\bea
 {\Psi_{L}}_A = U_{Ai}\psi_{Li}\;, \qquad {\Psi_{R}}_A =
V_{Ai}\psi_{Ri}\label{UpsiLR} \qquad \qquad
\mathbf{V^\dagger\, M\, U} = \mathbf{m} = \mathbf{diag}
(m_1,\ldots, m_N)
\eea
However, FET still applies using the following observation.  General
fermion propagator can be decomposed as
\bea
\frac{i}{\slashed k -\mathbf{M} P_L - \mathbf{M^\dag} P_R} 
= (\mathbf{M^\dag}P_L+\slashed{k} P_L) \frac{i}{ k^2 -\mathbf{M
    M^\dag}}+(\mathbf{M}P_R + \slashed{k} P_R) \frac{i}{ k^2
  -\mathbf{M^\dag M}}\label{mat-prop}
\eea
Therefore:
\begin{itemize}
\item Loop functions in fermionic amplitudes depend always only on
  eigenvalues of the Hermitian matrices {$\mathbf{M M^\dag}$} or
  {$\mathbf{M^\dag M}$}.
\item Only some combinations of mixing matrices can appear in
  fermionic amplitudes:
\bea
U_{Bi}\, f(m_i^2) \, U_{Ai}^{\star} &=& f(\mathbf{M^\dag
  M})_{BA}\nonumber\\
V_{Bi}\, f(m_i^2) \, V_{Ai}^{\star} &=& f(\mathbf{ M M^\dag
})_{BA}\nonumber\\
U_{Bi}\, m_i f(m_i^2) \, V_{Ai}^{\star} &=&{M}_{BC}^\dag\, f(\mathbf{M
  M^\dag})_{CA} = f(\mathbf{M^\dag M})_{BC} \, {M}_{CA}^\dag
\nonumber\\
V_{Bi}\, m_i f(m_i^2) \, U_{Ai}^{\star} &=& {M}_{BC}\, f(\mathbf{
  M^\dag M})_{CA} = f(\mathbf{ M M^\dag})_{BC} \, {M}_{CA}
\label{eq:fermionFET}
\eea
\end{itemize}
All such combinations can be expressed using FET formula for
$f(\mathbf{M M^\dag})$ or $f(\mathbf{ M^\dag M})$ , thus algebraic
expansion technique works for fermions as well. The conclusion holds
also for Majorana fermions -- then {${\bf M=M^T}$} and {${\bf
    U=V^*}$}.

\section{Automatized MI expansion: {\em MassToMI} Mathematica package}

FET technique has been automatized in the symbolic {\em Mathematica}
package {\em MassToMI}~\cite{Rosiek:2015jua}, able to expand mass
eigenstates amplitude to any user-defined MI order. Using {\em
  MassToMI}, one can:
\begin{itemize}
\item Calculate flavor amplitudes in the mass eigenstates basis, with
  the advantage of having less diagrams and more compact expressions,
  better suited for numerical computations.
\item Expand result using FET implemented in {\em MassToMI} package,
  recovering direct analytic dependence on ``interaction basis''
  parameters (better suited for understanding of various effects).
\end{itemize}

To illustrate {\em MassToMI} ~features, we consider as an example the
flavor violating charged Higgs boson decay $H^-\to d^I\bar u^J$ in the
MSSM ($I,J$ denote the quark generation indices). The sample diagram
contributing to the amplitude involving chargino (Dirac) {$C_n$},
neutralino (Majorana) {$N_j$} and down squark {$D_i$} circulating in
the loop is shown in the figure below:
\begin{center}
  \includegraphics[trim=9cm 205mm 75mm 53mm, clip=true,
  totalheight=0.2\textwidth]{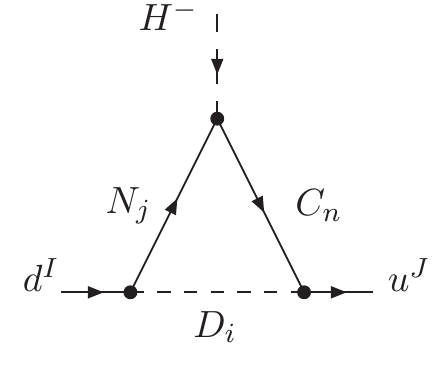}
\end{center}
For simplicity, again we choose just one of many terms in the
amplitude, having the ``typical'' form ({$Z,O,U,V$} denote scalar,
Majorana and Dirac fermion mixing matrices; summation convention is
assumed for the repeating indices):
\bea
A \supset Z_D^{Ii} Z_D^{Ji*} O_N^{Kj} O_N^{Lj*} V_C^{Mn*} U_C^{Nn}
m_{C_n} c_0(p,q,m_{C_n}^2,m_{D_i}^2,m_{N_j}^2)
\eea
To expand the amplitude, it has to be coded using the {\em MassToMI}
input syntax rules:\\[4mm]
\begin{tabular}{llll}
{$Z_D^{Ii}$} &$\to$ & {\small \tt {SMIX[D,I,i]}} & scalar $D$ mixing
matrix \\[1mm]
{$O_N^{Kj}$} &$\to$ & {\small \tt {NMIX[N,K,j]}} & Majorana fermion
$N$ mixing matrix\\[1mm]
{$V_C^{Mn*}$} &$\to$ & {\small \tt {Conjugate[FMIXL[C,M,n]]}} & left
Dirac fermion $C$ mixing matrix \\[1mm]
{$U_C^{Mn}$} &$\to$ & {\small \tt {FMIXR[C,M,n]}} & right Dirac
fermion $C$ mixing matrix\\[1mm]
{$m_{C_n}$} &$\to$ & {\small \tt {MASS[C,n]}} & physical fermion $C$
mass \\[1mm]
{$c_0(p,q,m_{C_n}^2,m_{D_i}^2,m_{N_j}^2)$} & $\to$ &
{\small \tt {LOOP[c0,\{\{C,n\},\{D,i\},\{N,j\}\},\{p,q\}]} } & loop integral \\
\end{tabular}
Then, {\em MassToMI} expression for the amplitude reads as:\\[2mm]
{\tt {
\begin{tabular}{ll}
A = & SMIX[D,I,i] Conjugate[SMIX[D,J,i]] \\
& NMIX[N,K,j] Conjugate[NMIX[N,L,j]] \\
& Conjugate[FMIXL[C,M,n]] FMIXR[C,N,n] MASS[C,n] \\
& LOOP[c0,\{\{C,n\},\{D,i\},\{N,j\}\},\{p,q\}];\\[3mm]
\end{tabular}
}}

\noindent Further, user must initialize control variables defining
order of MI expansion:
\begin{itemize}
\item {\tt FetScalarList = \{\{D,2\}\}}. Mixing matrices for scalar
  {$D$} are expanded up to the 2nd order in mass insertions.
\item {\tt FetFermionList = \{\{C,1,MHM\},\{N,1,MMH\}\}}.  Mixing
  matrices for fermions {$C,N$} are expanded to 1st order in
  MI. Parameters {{\tt MHM,MMH}} decide if the final result is
  expressed in terms of {$M^\dag M$} or {$M M^\dag$}.
\item {\tt FetMaxOrder = 2}.  Only mass insertion products of the
  total order {\tt FetMaxOrder} or lower are kept in the final result
  (e.g. terms proportional to 2nd order scalar MI and 1st order
  fermionic MI are truncated).
\end{itemize}
Using the setup above, command {{\tt \bf FetExpand[A] }} automatically
performs the MI expansion of amplitude to the required order. 

\smallskip

{\em MassToMI} package has been successfully tested on the realistic
case of the lepton flavor violating processes in the MSSM:
\smallskip
\begin{itemize}
\item Initial mass eigenstates expressions for amplitude: {few
  diagrams / few lines of code}.
\item Mathematica expansion code / execution time: {300 lines / up to
  few hours} on standard PC.
\item \textit{Intermediate expressions before cancellations: {$\sim
    50000$} MI terms!  Equivalent to evaluating of hundreds of
  diagrams in the interaction basis.}
\item Final MI expanded expressions after cancellations and
  simplifications: few lines for the leading MI terms.
\end{itemize}

\section{Conclusions}

I presented ``Flavor Expansion Theorem'', the general technique of
purely algebraic expansion of the mass eigenstates QFT amplitudes into
Mass Insertion series, valid for all types of amplitudes involving
scalar, fermionic or vector fields. Using FET allows to avoid tedious
and error prone direct diagrammatic calculations with MI's as
additional interaction vertices.  The technique can be fully
automatized with the use of specialized {\em MassToMI} package,
written in the {\em Mathematica} symbolic language.  The current {\em
  MassToMI} distribution, including detailed manual with examples, can
be downloaded from the address\\

\centerline{ {\tt www.fuw.edu.pl/masstomi}}

\Acknowledgements This work was supported in part by the National
Science Center, Poland, under research grants
DEC-2015/19/B/ST2/02848
and DEC-2015/18/M/ST2/00054.


\begin{thebibliography}{1}
  
\bibitem{Dedes:2015twa}
  A.~Dedes, M.~Paraskevas, J.~Rosiek, K.~Suxho and K.~Tamvakis,
  JHEP {\bf 1506} (2015) 151.

\bibitem{Rosiek:2015jua}
  J.~Rosiek,
  Comput.\ Phys.\ Commun.\  {\bf 201} (2016) 144.

\bibitem{Dedes:2014asa}
  A.~Dedes, M.~Paraskevas, J.~Rosiek, K.~Suxho and K.~Tamvakis,
  JHEP {\bf 1411} (2014) 137.

  
\end{thebibliography}
\end{document}